\documentstyle[prl,aps,preprint,tighten,floats,aps,amssymb]{revtex}

\newcommand{\scri}{\scriptsize}

\def\a{\alpha}
\def\b{\beta}

\def\g{\gamma}

\def\l{\lambda}

\def\o{\omega}
\def\p{\pi}

\def\J{\Psi}

\def\O{\Omega}

\def\bo{{\raise-.3ex\hbox{\large$\Box$}}}               
\def\face{{\raise.2ex\hbox{$\displaystyle \bigodot$}\mskip-2.2mu \llap {$\ddot
        \smile$}}}                                      
\def\tr{\mbox{\scri T}}                              

\def\ket#1{\left| #1\right\rangle}              
\def\leftrightarrowfill{$\mathsurround=0pt \mathord\leftarrow \mkern-6mu
        \cleaders\hbox{$\mkern-2mu \mathord- \mkern-2mu$}\hfill
        \mkern-6mu \mathord\rightarrow$}       
\def\dvec#1{\vbox{\ialign{##\crcr
        \leftrightarrowfill\crcr\noalign{\kern-1pt\nointerlineskip}
        $\hfil\displaystyle{#1}\hfil$\crcr}}}           

\def\beq{\begin{equation}}
\def\eeq{\end{equation}}

\def\beqx{\begin{displaymath}}
\def\eeqx{\end{displaymath}}

\def\beqa{\begin{eqnarray}}
\def\eeqa{\end{eqnarray}}

\begin{document}
\draft
\date{\today}
\title{
\normalsize
\mbox{ }\hspace{\fill}
\begin{minipage}{5cm}
UPR-890-T\\
{\tt hep-th/yymmxxx}
\end{minipage}\\[5ex]
{\large\bf D=4 N=1 Type IIB Orientifolds with 
Continuous Wilson Lines, Moving Branes, and their 
Field Theory
Realization\\[1ex]}}
\author{ Mirjam Cveti\v c  and 
Paul Langacker  }
\address{Department of Physics and Astronomy \\ 
University of Pennsylvania, Philadelphia PA 19104-6396, USA\\
}

\maketitle

\thispagestyle{empty}

\begin{abstract}
We  investigate four-dimensional N=1 Type IIB orientifolds with continuous
Wilson lines, and their T-dual realizations as orientifolds with moving
branes.  When  continuous Wilson lines  become discrete the gauge symmetry is
enhanced and  the T-dual orientifold corresponds to  branes sitting at  the
orbifold fixed points. There is a field theoretic analog describing these phenomena 
as  D- and F-flat deformations of the T-dual model, where the branes
sit at the origin (original model without Wilson lines)  as well as a
deformation of  the T-dual model where sets of branes sit at the fixed points (the
model with  discrete Wilson lines). We  demonstrate these
phenomena for the prototype $Z_3$ orientifold: 
we present an explicit construction of the   
general  set  of continuous Wilson lines  as well as their
 explicit field theoretic realization.
\end{abstract}
\newpage

\section{Introduction}
Four-dimensional N=1 supersymmetric Type IIB
orientifolds(see
\cite{ABPSS,berkooz,N1orientifolds,IbaneztypeI,TyeKakush,wilsonlinemodel,CPW,Aldazabal} and
references therein) provide a
domain of perturbative string vacua with novel properties (as opposed to the
perturbative heterotic solutions) with  potentially
interesting phenomenological implications. 
One goal, that is far from being achieved,  is the development of techniques that
would yield a larger class of solutions than those of based on symmetric
orientifold constructions.  However, even within the current fairly limited class
based on symmetric orbifolds, these  models possess a 
 rich structure of possible deformations  which  may provide a fruitful
ground for further investigation of their
phenomenological implications (with the ultimate goal to
identify classes of models with quasi-realistic features).

The deformations in the space of  supersymmetric four-dimensional solutions
  always
have a  field-theoretic realization, i.e.,
one identifies specific  D- and F-flat directions of the original
(undeformed) model. The new  (deformed) supersymmetric ground states 
correspond to  the exact string solutions, described  as a power-series
 in the magnitude of the vacuum expectation values of the  fields responsible
 for the deformations.
   In
particular, one interesting phenomenon to explore is the blowing-up of the
orientifold singularities \cite{CELW,douglas}, which are different in nature from
that of perturbative heterotic orbifolds \cite{C}.
This phenomenon is notoriously difficult to describe within the full string theory
context, since the metric of the blown-up space is not explicitly known.
On the other hand the explicit field-theoretic realization in terms of
(non-Abelian)  flat directions allows the determination of  the surviving
gauge groups and  massless spectrum.

Another set of deformations  corresponds to the 
introduction of  Wilson lines, both
continuous and discrete, and here one expects to have, parallel with the
field theory treatment, also the full string theory construction. 
The purpose of this paper is to address the study of such  continuous Wilson lines of
four-dimensional D=4 N=1 orientifolds, from both the full  string theory
description, i.e., by constructing explicitly  these Wilson
lines, and to find their T-dual interpretation,  as well as from the   
field theory side, i.e., by identifying the   moduli space of 
D- and F-flat directions of
the effective theory; these deformations correspond on the  
the  string side (in the T-dual picture) to the  ``motion'' of a set
of branes  away from the fixed points. However,  the
 construction of explicit  continuous Wilson lines 
 allows for  an  explicit string  theory  realization where 
sets  of branes  are located  at  an arbitrary distance  away from
 the fixed points.
On the other hand the field theoretical approach is only perturbative
 in the vacuum expectation values (VEV's), and thus in the string picture
corresponds only to a deformation
infinitesimally away from the undeformed model, i.e., where branes are located at
the orbifold fixed points.
 
The purpose of this paper is to set the stage for  constructions of 
 four-dimensional N=1 Type IIB orientifolds with continuous
Wilson lines, and their T-dual realizations as orientifolds with moving
branes.  In particular, we concentrate on explicit constructions of the
continuous Wilson line solutions and the corresponding field theory realizations
within the prototype, $Z_3$ orientifold model \cite{ABPSS}. 
The explicit realization of  these complementary pictures provide a beautiful
correpondence between the two approaches, and sets the stage for
further investigations of more involved orientifold models  with continuous
Wilson lines \cite{CLPUW}. 

Explicit examples of discrete Wilson lines have been constructed for 
a number of  a different orientifold models  
(see \cite{IbaneztypeI,wilsonlinemodel,CPW} and references therein).  
Continuous Wilson lines were first addressed in
\cite{IbaneztypeI}; however, the  explicit  unitary representation has not
been  given. The connection of models with
continuous Wilson lines  to the T-dual
models, where branes are located away from the orbifold fixed points, while
anticipated in general (\cite{lykken,ibanez,IbaneztypeI} and references
therein), was  exhibited  for a number of  examples
\cite{IbaneztypeI}.
It is also believed that in general 
there should be a field theoretical  realization of the same phenomena.

This paper advances these topics in several ways.  
In particular, we provide the first  explicit unitary representation of the
continuous Wilson line, specifically constructed  for the $Z_3$ orientifold,
and   construct for this model   the most  general set of continuous Wilson
 lines that in the
T-dual picture  correspond to the moving branes.
 We show that these solutions allow a continuous interpolation
between the original model without Wilson lines and the models with
discrete Wilson lines. 
In addition, we provide a systematic analysis of their  realization on the field theory side.

 The paper is organized in the following way. In Section IIa we summarize 
 the salient features of the  $Z_3$ orientifold construction. 
 In Section IIb we  proceed with the construction
 of, first discrete and then  continuous Wilson lines.
 When  continuous Wilson lines  become discrete the gauge symmetry is
enhanced and  the T-dual orientifold corresponds to  branes sitting at  the
orbifold fixed points. 
   In Section III we
 turn to the field theoretical analysis, by first
recapitulating  the techniques for  a  classification of D- and F-flat 
directions (Section IIIa). We  then   (Section IIIb) provide 
  explicit constructions of such  D- and F-flat 
directions and demonstrate their one-to-one correspondence with
 the continuous Wilson
line string constructions.

\section{$Z_3$ Orientifold  with Continuous Wilson Lines}
\subsection{ $D=4$, $N=1$  $Z_3$ orientifold}
We briefly  summarize the construction \cite{ABPSS} of  four-dimensional  $Z_3$
Type IIB orientifold models.  One starts with 
 Type IIB  string theory compactified  on a  $T^6/Z_3$  ($T^6$-six-torus,
 $G_1\equiv Z_3$ - the  discrete orbifold group)
  and mod out by the  world-sheet parity operation
$\O$, which is chosen to be accompanied by the same discrete symmetry
$G_2\equiv Z_3$ ,
i.e., the orientifold group is $G=G_1+\Omega G_2= Z_3 + \Omega Z_3$. 
 (Closure requires $\Omega g \Omega g'\in G_1=Z_3 $ for $g,g'\in G_2=Z_3$.)

The compactified tori are described by complex coordinates $X_i$,
$i=1,2,3$. The action of an orbifold group $Z_N$ on the compactified
dimensions can be summarized via  a twist vector $v=(v_1,v_2,v_3)$ (subject to
constraint $\sum _{i=1}^3 v_i=1$):

\beq 
  g: X_i \to e^{2i\pi v_i} X_i\;.
\eeq
For the $Z_3$ orientifold $v_1=v_1=v_3=\textstyle{1\over 3}$.

The tadpole cancellation, associated with the open-string modes, 
requires the inclusion of an even number of  $D9$ branes. (The case of
additional discrete symmetries in the orientifold group may require the
presence of multiple sets of $D5$ branes as well.)

An open string state is denoted as  as $\ket{\J,ij}$ where $\J$ denotes the 
world-sheet state and $i,j$ the Chan-Paton indices associated with the end
points on a $D9$ brane. The elements $g\in G_1=Z_3$ act on open string 
states as follows:
\beq
  g: \ket{\J,ij} \to 
  (\g_g)_{ii'}\ket{g\cdot\J,i'j'}(\g_g^{-1})_{j'j}\;.
  \label{ori1}
\eeq
Similarly, the elements of $\O G_1=\O Z_3$ act as
\beq
  \O g: \ket{\J,ij} \to (\g_{\O g})_{ii'}\ket{\O g\cdot\J,j'i'}
  (\g_{\O g}^{-1})_{j'j}\;,
  \label{ori2}
\eeq
where we have defined $\g_{\O g}=\g_g\g_{\O}$, up to a phase, in
accordance with the usual rules for multiplication of group elements.
Note, that $\O$ exchanges the Chan-Paton indices.

Since the $\g_g$ form a projective representation of the orientifold
group, consistency with group multiplication implies some conditions
on the $\g_g$. Consider the $G_1=G_2=Z_3$ case;  $g^3=1$, $\O^2=1$ and
$\O g \O g' \in Z_3$ for $g,g'\in Z_3$ respectively imply:
\beq
 \g_g^3=\pm 1\; \ \ , 
  \g_{\O}=\pm \g_{\O}^{\tr}\;\ \ , 
  (\g_g^k)^*=\pm\g_{\O}^*\g_g^k\g_{\O}\;.
  \label{gstar}
\eeq
It turns out that the tadpoles cancel, if we choose the plus sign for
$D9$  branes (and the minus sign for $D5$ branes) \cite{IbaneztypeI}.
The explicit representation for  the $D9$ brane sector $\g_\O$ is
symmetric and can be chosen real:
\beq
  \g_{\O,9}=\left(\begin{array}{cc}0&\openone_{16}\\ 
  \openone_{16}&0\end{array}\right)
\;,
  \label{omega}
\eeq
where the subscript $9$ 
denotes the $D9$ brane sector in which these
matrices are acting.

Further, finiteness of string loop diagrams yields tadpole
cancellation conditions which constrain the traces of $\g_g$ matrices
\cite{ABPSS}
\begin{equation}
\label{z3trace}
{\rm Tr}(\gamma_{Z_3}) = -4\;. 
\end{equation}

The $Z_3$ twist action on the tori is given by the twist vector
$v=({1\over 3},{1\over 3},{1\over 3})$ and
its action on Chan-Paton matrices is
generated by:
\beq
  \g_{Z_3}=\mbox{diag}\left(\o\openone_{12},\openone_4,
           \o^2\openone_{12},\openone_4\right)\;,
  \quad\mbox{where}\quad\o=\mbox{e}^{2\p i/3}\;. \label{gammaZ3}
\eeq

This choice satisfies eqs.~(\ref{gstar}) and (\ref{z3trace}).   
Open string states, whose Chan-Paton matrices will be denoted by
$\l^{(i)}$, $i=0,\ldots,3$ in the following, give rise to space-time
gauge bosons ($i=0$) and matter states ($i=1,2,3$). 

 Gauge bosons in
the $D9$ brane sector arise from open strings beginning and ending on
$D9$ branes.  Invariance of these states under the action of the
orientifold group requires
\beq
  \l^{(0)}=-\g_{\O,9}{\l^{(0)}}^{\tr}\g_{\O,9}^{-1} 
  \qquad\mbox{and}\qquad
  \l^{(0)}=\g_{g,9}\l^{(0)}\g_{g,9}^{-1}\;.
  \label{cond1}
\eeq
With eq.~(\ref{omega}) the first constraint implies that the
$\l^{(0)}$ are SO(32) generators, while the constraints from the
$\g_{g,9}$ will further reduce the group.

The result is the gauge group:
\beq
 U(12)\times SO(8)\; .
\eeq

The Chan-Paton matrices of the matter
states have to be invariant under the action of the orientifold
group as well. However, since the string vertices for the 
 chiral matter superfields involve the oscillator modes of the target
 space toroidal coordinates $X_i$, the  Chan-Paton matrices now transform
under
 the orbifold action (in order to render the  physical  states invariant
 under the orbifold action), thus implying:
 
\beq
  \l^{(i)}=-\g_{\O,9}{\l^{(i)}}^{\tr}\g_{\O,9}^{-1} 
  \qquad\mbox{and}\qquad
  \l^{(i)}=\mbox{e}^{2i\p v_i}\g_{g,9}\l^{(i)}\g_{g,9}^{-1}\;.
\eeq

For $Z_3$ orientifold this yields   a matter content of three copies of
\beq\psi^\alpha=(\overline{12},8)_{-1}, \ \ \
 \chi^\alpha=({66},1)_{+2} \ \ , \  \
\alpha=1,2,3.
\eeq
where the subscript refers to the $U(1)$ charge of U(12).
 The closed string sector  yields the
gravity supermultiplet and the 36 (chiral) supermultiplets corresponding to
the $9$ untwisted (``toroidal'') and $27$ twisted (blowing-up)  sector moduli.
The moduli  are   gauge singlets, whose real and imaginary components
arise from the NS-NS and R-R sector, respectively.

The renormalizable superpotential is of the form
\begin{equation}
{\cal W}\sim \epsilon_{\alpha \beta \gamma} \psi^{\alpha a}_{i} \psi^{\beta b}_{i}
\chi^\gamma_{[a,b]},
\label{Weqn}
\end{equation}
where  $\alpha, \beta, \gamma$ are family indices, $\{a,b\}$-$U(12)$ indices, 
and $i$-$SO(8)$ indices.
\subsection{Wilson Lines}

\noindent{\bf Discrete Wilson Line}\ \  When the action of the Wilson line 
   on the Chan-Paton matrices, which is
  represented by  a matrix 
$\gamma_W$,    is  such that it commutes  with $\gamma_{Z_3}$,  it 
 depends only on  discrete values of parameters, i.e., it   describes a
discrete Wilson line.   Let us  focus on  a Wilson line, acting along 
the two-torus  coordinate $X_i$, say $i=3$.   
It satisfies 
the following algebraic consistency conditions:
\beq
  \left(\g_{Z_3}\g_W\right)^3=+1\  , \ [\gamma_{Z_3}, \gamma_W]=0\ .
  \label{Wcond1}
\eeq

 Further, tadpole cancellations require
\beq
 \mbox{Tr}\left(\g_{Z_3}\right)=\mbox{Tr}\left(\g_{Z_3}\g_W\right)
  =\mbox{Tr}\left(\g_{Z_3}\g_W^2\right)=-4\; ,
  \label{Wcond3}
\eeq
To simplify the notation, let us rearrange  the entries in the
$\g_{Z_3}$ matrix  (\ref{gammaZ3}) as follows:
\beqa
  \g_{Z_3}&=&\mbox{diag}\left(\o^2\openone_{6-n},\o
  \openone_{6-n},\openone_{4-n},   {\bf Z}\otimes\openone_{n};
  \o\openone_{6-n},\o^2
  \openone_{6-n},\openone_{4-n},   {\bf Z}^*\otimes\openone_{n} \right)\;;\\[1ex]
  {\bf Z}&=&\mbox{diag}\left(\o^2,\o,1\right) ,\quad\  
 \o=\mbox{e}^{2\p i/3}\;, \ n=\{0,\cdots,4\} \; . \label{gammaZ3P}
\eeqa
The above consistency conditions  reduce to the unique 
 solution of the discrete Wilson line:

\beq
  \g_W=\mbox{diag}\left(\openone_{16-3n}, 
    \o\openone_{3n};
    \openone_{16-3n},\o^2\openone_{3n}\right)\; . 
     \label{wilson}
\eeq

The surviving gauge symmetry 
is determined by a projection:
\beq
  \l^{(0)}=\g_W\l^{(0)}\g_W^{-1}\;,
\eeq
which further breaks the gauge group down to:

\beq
U(12-2n)\times SO(8-2n) \times  U(n)^3\; .
\label{discgg}\eeq

The matter representation is determined by the condition:
\beq
  \l^{(i)}=
\g_W\l^{(i)}\g_W^{-1}\;.
\eeq
The matter comes in three copies  and has the following representation (the
subscripts correspond to  the self-evident $U(1)$ charges):

\beqa
\chi^\a&=&((6-n)(11-2n), 1,1,1)_{-2}\; ,  \label{disc1} \\
\psi^\a&=&(\overline{12-2n},8-2n,1,1,1)_{1}\; ,\\
S^\a&=&(1,1,n,1, \bar{n})_{1,-1}\; ,\\
P^\a&=&(1,1,\bar{n},n,1)_{-1,1}\; ,\\
Q^\a&=&(1,1,1,\bar{n},n)_{-1,1}\; ,
\ \ \a=1,2,3\;
.\label{discmatter}\eeqa
The superpotential is of the form:
\beq
{\cal W}\sim \epsilon_{\alpha \beta \gamma} \psi^{\alpha a}_{i} 
\psi^{\beta b}_{i}\chi^\g_{[a,b]}
+S^{\a{ i_3}}_{i_1} P^{\b{ i_1}}_{i_2} 
Q^{\gamma{ i_2}}_{i_3}
\; , \ \ \ 
\rm{with} \;  \a\ne\b\ne\gamma, 
\eeq
where  $\alpha, \beta, \gamma$ are  again family indices,
 $\{a,b\}$-$U(12-2n)$ indices, 
$i$-$SO(8-2n)$ indices, and  
$\{i_{1,2,3}\}$-respective indices for
the three $U(n)$ factors.

There is a T-dual interpretation of this solution.  Namely, T-dualizing the
original model along, say,   the  third  complex direction $X_3$,
 corresponds to the 
model with 32  $D7$ branes sitting at  the origin of  the third complex
 plane.  The action of  discrete Wilson lines implies that  one can take
  $n$ sets  ($n=1,\cdots, 4$) of branes (each set
 containing  six $D7$ branes)  to be 
 at one of the  two  $Z_3$ orbifold fixed points, 
 located away from  the
 origin. 
[This motion  can be accomplished in sets of six $D7$  branes;
of  six
since each such set is 
moded out  by
   six elements of
 the combined $Z_3$ and $\O$ group action.]
Note that the string states associated with branes  located at one  $Z_3$
orbifold fixed point, away from the origin (in a particular complex
plane), are
related, by  orientifold projection,  to complex conjugate states of
branes located at the other fixed point  away from the origin.

\vskip 0.5cm

\noindent{\bf Continuous Wilson Lines}\ \ 
As the next step  we generalize the construction to the case of 
continuous Wilson lines.  We are still after a unitary representation $\g_W$ that
satisfies conditions (\ref{Wcond1})-(\ref{Wcond3}), except that 
  now it need not commute with $\g_{Z_3}$.
  
For convenience, we again rearrange the entries
in   the $\g_{Z_3}$ matrix,
\beq
  \g_{Z_3} = \mbox{diag}\left(\o^2\openone_{6-n},\o
  \openone_{6-n},\openone_{4-n}, \openone_n\otimes  {\bf Z};
  \o\openone_{6-n},\o^2
  \openone_{6-n},\openone_{4-n},  \openone_n\otimes {\bf Z}^* \right)\;.
\label{gammaZ3PP}
\eeq
%
The Ansatz for the Wilson line  is taken to be of the  form:
\beq
\gamma_W=\mbox{diag}\left(\openone_{16-3n},\g_W^0;
\openone_{16-3n},(\g_W^0)^*\right),
\eeq
where $\gamma_W^0$ is a $(3n\times 3n)$-dimensional unitary matrix.
Such a  Wilson line can be  uniquely determined up to
unitary transformations that commute with $\g_{Z_3}$. The part of
such unitary transformations that affects $\gamma_W^0$   is of the form:
$
U_n\otimes U_3^0$, where
$U_n$ is a general $(n\times n)$-dimensional unitary matrix and  $U_3^0$
is a $
(3\times 3)$-dimensional diagonal unitary matrix. Employing such
transformations in turn enables one to cast $\gamma_W$ in the following most
general form:
\beq
  \g_W=\mbox{diag}\left(\openone_{16-3n}, 
    W_1,\cdots, W_n;\openone_{16-3n},
    W_1^*,\cdots, W_n^*\right)\ , \label{WL}
    \eeq
  where  $W_i$ are ($3\times 3$)-dimensional  unitary matrices subject to
the following
    consistency conditions:
    \beq
   {\rm Tr}({\bf Z}W_i)= {\rm Tr}({\bf Z} W_i^2)=0\; , \ \ ({\bf Z}W_i)^3=
\openone_3\;
   .\label{trace}
   \eeq
   
We have reduced the  problem to finding  an explicit representation of the
  matrices $W_i$. Starting with a general Ansatz:
  \beq
W= \left(\begin{array}{ccc}w_1&a&b\\ 
  a'&w_2&c\\
  b'&c'&w_3\end{array}\right)\ \; ,
  \eeq
the  conditions (\ref{trace})
 yield the following general form:
  \beq
W_0= \left(\begin{array}{ccc}w&a&b\\ 
  a'&w+x&c\\
  {{aa'+\o wx}\over b}& {{aa'-\o^2 x(x+w)}\over c}&w-\o^2
x\end{array}\right)\ \; , 
 \label{mat} \eeq
where $a,b,c,w,x$  are complex numbers subject to the
constraint ${\rm det}\
(W)=1$. However,  unitarity imposes  an immediate constraint $x=0$; this
is due to the fact that
 the  co-factors  of the diagonal entries  in  the above matrix   are all
  equal to $w(w+x)-aa'$, and  due to unitarity they should be proportional
  to the complex conjugate  diagonal  entries  in (\ref{mat}).

 The  unitarity conditions (equating $W^\dagger$ entries with the corresponding
 co-factors of $W$) allow one to further constrain the parameters:

\beq
|c|=|a|,\ \  |b|=|a'|,\ \  |w|=\sqrt{1-|a|^2-|a'|^2}\ ,  
\eeq

At this point, for the sake of simplicity, 
 we shall change the notation $(|a|,|a'|,|w|) \to (a,a',w)\ge 0$,
subject to the constraint $w=\sqrt{1-a^2-a'^2}$.

To further  determine the
phases $\phi_a,\phi_b,\phi_c,\phi_{a'}, \phi_w$, 
we introduce
\beqa
\phi_A&=&2\phi_{a'}+\phi_a+\phi_b-\phi_c\; ,\\
\phi_B&=&2\phi_a +\phi_{a'}-\phi_b+\phi_c\; ,\\
\phi&=&\phi_w+\phi_a+\phi_{a'}\; ,
\eeqa
and reduce the remaining unitarity conditions 
 to the following set of  equations:
\beqa
 -a a' w e^{i\phi} + a'^3 e^{i\phi_A} &=&a'^2 \; ,\label{phase1}\\
 -a a' w e^{i\phi} + a^3  e^{i\phi_B}& =& a^2 \; ,\label{phase2}\\
 -a a' w e^{i\phi} + w^3 e^{3i\phi_w}& =& w^2 \; .
\label{phase3}\eeqa
[$({\bf Z}W)^3=\openone_3$ is automatically satisfied, provided 
(\ref{phase1})-(\ref{phase3})  
hold;  namely, (\ref{phase1})+(\ref{phase2})+(\ref{phase3})
 is precisely the  condition ${\rm det} \ ( W)=1$.]
The solution of  (\ref{phase1})-(\ref{phase3}) gives:
\beqa
\cos(\phi_A)&=&{{(a^4+a^2a'^2+a'^4+a'^2-a^2)}\over{2a'^3}}\;,\label{cos1}\\
\cos(\phi_B)&=&{{(a^4+a^2a'^2+a'^4-a'^2+a^2)}\over{2a^3}}\; , \label{cos2}\\
\cos(3\phi_w)&=& {{(w^4+w^2-a^2a'^2)}\over{2w^3}}\; , \label{cos3}
\eeqa
as well as 
\beq
\cos(\phi)  ={{(a^4+a^2a'^2+a'^4-a'^2-a^2)}\over{2a a' w}}\; .
\label{cos4}\eeq

Two additional constraints  of eqs. (\ref{phase1})-(\ref{phase3}) turn out to
 be automatically satisfied. In addition, 
the  phases  $\phi_A$, $\phi_B$, $\phi_w$ and $\phi$ 
are not independent, i.e., 
\beq
3\phi= 3\phi_w + \phi_A + \phi_B.  \label{constraint}
\eeq
One can show   that the expressions (\ref{cos1})-(\ref{cos4}) 
indeed  ensure $\cos(3\phi)=\cos(3\phi_w + \phi_A + \phi_B)$.
(In proving this identity, the
relationships: $a'^3 \sin(\phi_A)=a^3 \sin(\phi_B)=w^3\sin(3\phi_w)=a\,a'\,
w\sin(\phi)$ that follow from (\ref{phase1})-(\ref{phase3}) are useful.)

Thus we have arrived at the following form of the  matrix $W$:
  \beq
W= \left(\begin{array}{ccc}w\,e^{i\phi_w}&a\,e^{i\phi_a}&a'e^{i\phi_b}\\ 
  a'\,e^{i\phi_{a'}}&w\,e^{i\phi_w}&a\,e^{i\phi_c}\\
  a\,e^{i(\phi_a+\phi_{a'}-\phi_b)}&a'\, e^{i(\phi_a+\phi_{a'}-\phi_c)}&w\,e^{i\phi_w}
  \end{array}\right)\ \;. \label{WL0}
  \eeq
It is  determined  up to  diagonal unitary
transformations $U_3^0={\rm diag}(e^{i\varphi_1},e^{i\varphi_2},e^{i\varphi_3})$,
which is the most general unitary matrix that commutes with ${\bf Z}$, and thus
provides an equivalency
 class for the Wilson line representations. Thus,
two phase parameters in $W$  can be ``gauged away'', and  the only
remaining phases are  the two ``gauge invariant'' phase parameters $\phi_{A,B}$
and the phase of the diagonal element $\phi_w$, all  specified,
 by eqs.
  (\ref{cos1})-(\ref{cos3}), in terms of two real parameters,
up to signs and multiples of $2 \pi$. The latter are further constrained
by (\ref{phase1})-(\ref{phase3}) and (\ref{constraint}).

To summarize,  the final form of (\ref{WL0}) is  
specified in terms of three  real, positive parameters $w,a,a'$, 
subject to the constraint $w^2+a^2+a'^2=1$. [Equivalently the 
Wilson line can be  specified in terms of 
the two Euler angles  $\varphi$ and $\psi$ of the three-sphere,
introduced as:
$w=\cos\psi, \ \  a=\sin\psi\cos\varphi, \ \ 
a'=\sin\psi\sin\varphi$.]

To study the monodromy properties of this Wilson line, it is convenient
to solve  (\ref{cos3}) for $\cos{\phi_w}$, which has three roots:

\beqa
\cos(\phi_w)_{1}&=& \textstyle{1\over 2}\left(e^{i\phi_0}+e^{-i\phi_0}\right)\;
, \\
\cos(\phi_w)_{2}&=& \textstyle{1\over 2}\left(
                      \omega e^{i\phi_0}+\omega^2 e^{-i\phi_0}\right)\; ,\\
\cos(\phi_w)_{3}&=&\textstyle{1\over 2}\left(   
                 \omega^2 e^{i\phi_0}+\omega e^{-i\phi_0}\right)\; ,
\eeqa

where 
\beq
e^{i\phi_0}\equiv (A+i\sqrt{1-A^2})^{1/3} \; , \ \ 
A\equiv{{(w^4+w^2-a^2a'^2)}\over{2w^3}}\; . 
\eeq

For the limit $A=1$, i.e., $w=1,\ a=a'=0$, these solutions reduce to
$\phi_w=0, \omega, \omega^2$, which are respectively the
case of no Wilson line, or discrete Wilson lines,
corresponding (in the T-dual picture) to the set of  $D7$ branes sitting
at the origin or at
one of the two fixed points. The continuous Wilson line interpolates
continuously between these limits.
For that purpose one has to 
find a path  in the space of  $w,a,a'$ (or equivalently the Euler 
 angles  $\varphi$ and $\psi$)  
in which $3\phi_0$ varies from $0$ to  
$\pm2\pi$.
 It is straightforward to find such paths. In particular,
$3\phi_0=0$ and $3\phi_0 =\pi$ correspond to $A=+1$ and $-1$,
respectively, which occur on the boundaries of the allowed
region at the points $a=a'=0$ and $a=a'=\frac{2}{3}$.
 There is a continuous path between these,
which passes $3\phi_0 ={\pi\over 2}$ ($A=0$) at, e.g., $a = a' =
1-\frac{1}{\sqrt{3}}$. From $A=\pm 1$ one can always move along
either branch of the square root, allowing a
 continuous interpolation, e.g.,  from  $3\phi_0=0$ to
  $3\phi_0 =\pi$ and then to $3\phi_0=2\pi$ and further.
For example, $3\phi_w$ can start from $0$ and increase first to $2\pi$
and then to $4\pi$ as one moves back and forth between $A=+1$ and $A=-1$.
An examination of (\ref{phase1})-(\ref{constraint}) reveals that
$\phi_A$ and $\phi_B$ each decrease by $\pi$ as $3\phi_w$
increases by $2\pi$. As $3\phi_w$ passes through $2\pi$, $\phi_{A,B}$ and $\phi$
must increase discontinuously by $3\pi$ and $2\pi$, respectively, in order
to preserve the signs of the angles. This occurs at $a = a' =0$,  where
the phases are indeterminant, so the changes in the Wilson line
parameters are continuous. It is convenient to use the freedom
in $U^0_3$ to require $\phi_b=\phi_c$, in which case an increase
of $\phi_w$ by ${{2\pi}\over 3}$ (or ${{4\pi}\over 3}$) is accompanied by
an increase
in $\phi_{a,a'}$ by the same amount. From (\ref{trace}) it is clear that if
$W$ is a continuous Wilson line solution, then so are $\o W$ and $\o^2 W$. We thus 
see that these solutions, related by the discrete $Z_3$ symmetry
for fixed $a$ and $a'$, can 
actually be related by a continuous interpolation as $3\phi_w$
varies from $0$ to $2\pi$ to $4\pi$ as a function of $a$ and $a'$.

The surviving gauge group is generically:
\beq
U(12-2n)\times SO(8-2n)\times U(1)^n\;, \label{gaugegr}
\eeq
as long as   $W_1\ne W_2\ne  \cdots  W_n$ and  $(a_i,a'_i)\ne 0$
($i=1,\cdots ,n$). However for special values of the $W_i$ parameters
additional gauge
enhancement can take place. In particular, $W_1=W_2\dots =W_k$ ($k\le n$) 
yields the gauge group enhancement of $U(1)^n$ to $U(k)\times
U(1)^{n-k}$,
 with an obvious generalization to two sets $W_1=W_2$ and $W_3=W_4$.

This general continuous Wilson line reduces to a  hybrid (continuous/discrete)
Wilson line if $k$ $W$ matrices become  diagonal with elements  $\omega$
(discrete Wilson lines) 
(or equivalently, $\omega^2$) and the remaining ($n-k$) $W$-matrices remain
off-diagonal   (continuous). The gauge group  $U(1)^n$  now 
becomes enhanced to  $U(k)^3\times U(1)^{n-k}$.

The Wilson line (\ref{WL})  has a T-dual interpretation in terms of
$n$ sets of  six $D7$ branes each  moving in, say, 
 the third complex plane away from
the  orbifold fixed  at the origin. In particular, when a subset of $k$  Wilson line 
 elements
become discrete, the T-dual picture  corresponds to
  $k$-sets of six $D7$ branes   sitting at one (of the two)
  orbifold fixed point away from the origin.

The above construction of the   Wilson lines  for the  $Z_3$ orientifold
provides a generalization of the discrete Wilson lines (with 
 $n=4$)  discussed in \cite{IbaneztypeI}. The continuous
Wilson line considered  there  corresponds to the case with
$W_1=W_2=W_3=W_4$. Here, we have generalized and given an
 explicit unitary representation of $W_i$ in terms
  of two parameters.  
(When one imposes the unitarity constraint on the symmetric
matrix in \cite{IbaneztypeI}, 
there is a relation between the magnitude and phase of
the complex parameters, and the matrix
becomes a special  one parameter case  (with
$a=a'$) of the continuous Wilson described above.)
In the  T-dual picture the   position  of the branes (in the
third complex plane) should be parameterized by two real parameters, and thus 
for the full one-to-one correspondence with the continuous Wilson line picture
the Wilson line should depend on two real parameters ($a$ and $a'$) as
well.
 
\vskip 0.5cm
\noindent{\bf Multiple Continuous Wilson Lines}\ \ 
 As the last step we proceed with the construction of multiple Wilson lines. 
 On general grounds such Wilson lines  are of the form (\ref{WL})
 with the $W_i$'s being of the form (\ref{WL0}). The elements in (\ref{WL0})
 are  uniquely specified  in terms of  two real  parameters, except for two
 phases. As discussed above, for a single Wilson line these two phases  can be
 removed  (``gauged away'') by a diagonal unitary transformation $U_3^0$ that
 commutes with $\bf Z$. 
 
 For the  additional Wilson lines $\gamma_{W_J}$ one cannot gauge away the two
 phases. However,  the Wilson lines have to commute:
 \beq
[\g_{W}, \g_{W_J}]=0\; .
\eeq 
This condition should in principle fix the
undetermined phases. We checked that this is indeed the case. 
We chose  the  first Wilson line $\gamma_{W}$ 
by fixing the gauge, i.e., choosing a specific $U_3^0$, so that
 the phases for the matrices $W$ are fixed as:
\beq
\phi_a=\phi_{a'}=\phi_b\; .
\eeq 
Then, if  the phases for the corresponding matrices $W_{II}$ in 
the  second Wilson line $\gamma_{W_{II}}$ 
are chosen as
\beq
\phi_{c_{II}}=\phi_{a_{II}}+\phi_c-\phi_a\; , \ \ 
\phi_{a'_{II}}=\phi_{b_{II}}\; ,
\eeq
the two Wilson lines  commute. 

One can  introduce the third Wilson line $\g_{W_{III}}$, with the
 phases subject to the analogous constraint:
\beq
\phi_{c_{III}}=\phi_{a_{III}}+\phi_c-\phi_a\; , \ \ 
\phi_{a'_{III}}=\phi_{b_{III}}\; .
\eeq 
Such a Wilson line turns out to commute both with $\gamma_{W}$ 
and $\gamma_{W_{II}}$.  (Of course, one still has the
freedom to make an overall unitary transformation $U_3^0$ on
all three Wilson lines.)

The additional Wilson line $\gamma_{W_{II}}$   has again a 
  T-dual interpretation.   (The introduction 
  of additional Wilson lines does not break the
generic gauge group (\ref{gaugegr}).)
  T-dualizing the
original model along two, say,  the  second $X_2$ and third
$X_3$,  complex directions corresponds to the model
with   32  $D5$ branes sitting at  the origin of  the second and  third complex
 plane.  
 The action of  continuous Wilson lines  $\gamma_{W_{II}}$  and $\g_W$
  then corresponds to the independent motion 
   of $n$ sets of six $D5$ branes 
 in the second  and the third complex planes, respectively. Since each $W$  and
 $W_{II}$ are fully specified by two real parameters, these parameters
  are in one-to-one
 correspondence with the motion  of the ($n$) sets  of  branes 
 in the two complex planes.
 Again,   when any Wilson line element becomes diagonal (discrete) the
 solution corresponds to a particular set of branes reaching the orbifold
fixed  point in the particular plane.

  The T-dual model, in which one has dualized all 
 three complex directions $X_{1,2,3}$, corresponds to the $C^3/Z_3$ model of 32
 $D3$ branes sitting at the origin.  The introduction
  of the third Wilson line $\gamma_{W_{III}}$, which is   also uniquely
 specified by two  real prameters for each $W_{III}$ matrix,
   parameterizes the  independent
  motion of $n$ sets of (six) $D3$ branes along the first plane (along
   with the independent motions in the second
  and third planes,  parameterized by $\g_{W_{II}}$ and $\g_W$, respectively).

 \section{Field Theory Realization}

\subsection{Classification of F- and D-flat directions}

In this section we classify the
F- and D-flat directions of the 
new supersymmetric ground states that  correspond to the deformation of the
original model (and whose string theoretical construction we provided  in the
previous Section). We also show how to deform
from the discrete Wilson line solutions corresponding to $n$ sets of branes
located at the orbifold fixed points away from the origin.
For that purpose we utilize
 the one-to-one correspondence  \cite{monomialpapers} of D-flat directions
with
holomorphic gauge-invariant polynomials (HIP's) built out of the chiral
fields in the model. The
constraints of F-flatness 
further require that $\langle \partial W /
\partial \Phi_{p} \rangle =0$ and $\langle W \rangle =0$ for all of 
the massless superfields $\Phi_{p}$ in the model.
(The detailed analysis of the blown-up $Z_3$ orientifold was given in
\cite{CELW}, using this technique\footnote{These techniques were 
previously developed
 (see \cite{cceel2}) to construct the moduli space of
  the flat directions for models based on perturbative heterotic string
  models.
For simplicity,
the flat direction analysis in \cite{cceel2} considered only the
non-Abelian
singlet fields in the model, in which case the flat directions correspond
to gauge invariant holomorphic
monomials.
In the present model, the D-flat directions necessarily involve
non-Abelian fields due to the matter content. }. )

We first construct a gauge invariant polynomial from the
non-Abelian fields, which is a sum of monomials involving the components
of the fields. Then one monomial term defines a D-flat direction. 
Each field
in the monomial will typically have the same  vacuum expectation
value (VEV). The D-flat constraints for
both diagonal and off
diagonal generators of the non-Abelian gauge group are automatically   
satisfied. Other flat directions, e.g., those with different phases for
the VEV's of  the fields
in the monomial,  are gauge rotations of the original monomial. 

One can also consider  D-flat directions
with more than one independent VEV, formed as products of
other HIP's. The flat directions correspond to products
of monomials from each of the HIP's, each with its own VEV. (See
\cite{CELW} for more detailed discussion of such issues as overlapping
HIP's, involving products of HIP's which have common multiplets. Here,
it suffices to check D-flatness for each case.)

\subsection{Flat Directions corresponding to Wilson Line Solutions}

Unlike the blowing-up procedure \cite{CELW,douglas} in which the  
blow-up introduced 
Fayet-Iliopoulos terms \cite{douglasmoore} for the anomalous $U(1)$ terms,  and thus the
supersymmetric ground state solutions were achieved by HIP's which have {
non-zero} $U(1)$ charges,  the deformations corresponding to the continuous
Wilson lines correspond to the polynomials that  are  gauge invariant under
the ``anomalous'' $U(1)$ as well.

The D-flatness condition for $SO(8)$ is
\beq
D^{I}=\sum_{\alpha, a}\sum_{i,j}(\psi_{i}^{\alpha a \dagger} A^{I}_{ij}
\psi_{j}^{\alpha a })=0,
\eeq
where $A^{I}$ are generators of the vector representation of $SO(8)$ and
$I=1,..,28$. For $U(12)$,

\beq D^J = \sum_{\alpha,i} \sum_{a,b} \psi_i^{\alpha a \dagger}
\hat{T}_{ab}^J \psi_i^{\alpha b}
+ \sum_{\alpha} \sum_{a,b,c} \left( \chi^\alpha_{[a,c]}\right)^\dagger
T^J_{ab} \chi^\alpha_{[b,c]},  \eeq
where $T^J$ ($\hat{T}^J \equiv - T^{JT}$) are the generator matrices for
the fundamental (anti-fundamental) representation of
$U(12)$ and $J=1,..,144$. (In the blown-up case one must
add a constant Fayet-Iliopoulos term $\xi_{FI}$ to the D-term
for the
anomalous $U(1)$ \cite{CELW}.)

The F-flatness conditions of the original $Z_3$ orientifold  model are
\beq
\epsilon_{\alpha \beta \gamma} \psi_{i}^{\alpha a} \psi_{i}^{\beta b}=0; \ \
\ \
\epsilon_{\alpha \beta \gamma} \psi_{i}^{\beta b} \chi^{\gamma}_{[a,b]}=0.
\eeq

For the case of the original $Z_3$ orientifold, one can construct D- and
F-flat directions
from HIP's  of the
form $(\psi \psi \chi) (\psi \psi \chi)$, in which  each
factor is
separately $U(12)$ invariant, and only the product is
$SO(8)$, invariant, i.e.,
\beq
 (\psi^{\alpha a}_i \psi^{\alpha b}_j \chi^{\alpha}_{[a,b]}) 
(\psi^{\alpha c}_i \psi^{\alpha d}_j \chi^{\alpha}_{[c,d]}). \label{basic}
\eeq
F-flatness is ensured
by taking all fields from the same family, e.g., $\alpha =3$ for definiteness.
It is necessary to consider a 6$^{th}$ order polynomial 
because the cubic $(\psi \psi \chi)$ vanishes for a single
family index due to the symmetry [antisymmetry] in $SO(8)$ [$U(12)$]
indices.
A flat direction corresponds to  a specific monomial in (\ref{basic}),
e.g., to
\beq
 (\psi^{3 1}_1 \psi^{3 2}_2 \chi^{3}_{[1,2]}) 
(\psi^{3 1}_1 \psi^{3 2}_2 \chi^{3}_{[1,2]}). \label{flat}
\eeq
This is just the square\footnote{It is easy to check that
(\ref{basic}) is only D-flat for $[a,b]=[c,d]$. For example,
$(\psi^{3 1}_1 \psi^{3 2}_2 \chi^{3}_{[1,2]}) 
(\psi^{3 3}_1 \psi^{3 4}_2 \chi^{3}_{[3,4]})$ is not D-flat under
$U(12)$ because, in the terminology of
\cite{CELW}, it involves overlapping $U(12)$ polynomials.
} of $\psi^{3
1}_1
\psi^{3 2}_2
\chi^{3}_{[1,2]}$, with each of the three fields having the common VEV $v$,
and will henceforth be denoted by $(\psi^{3
1}_1
\psi^{3 2}_2
\chi^{3}_{[1,2]})^2$.

The direction (\ref{flat}) appears to break $SO(8) \times U(12)$ to
$SO(6) \times U(10)$. In fact, there is an additional surviving
$U(1)$ generated by a combination of the broken $SO(8)$ and $U(12)$
generators. To see this, consider a
concrete representation of
non-Hermitian $SO(8)$ and $U(12)$ generators
labeled by $I=(kl)$ and $J=(cd)$, i.e., 
 $A_{kl}$ and $T^c_d$, with
matrix elements
\beq
(A_{kl})_{ij} = \delta_{jk}\delta_{il}-\delta_{jl}\delta_{ik},
\eeq
and 
\beq
(T^c_d)_{ab} =  \delta_{ad} \delta_{bc},
\eeq
respectively. It is then straightforward to see that
the Hermitian generator $F_{12}$, defined by
\beq
F_{ij} = i\left(T^i_j - T^j_i + A_{ij}\right),
\eeq
is conserved.

Several generalizations of (\ref{flat}) are possible. One
can consider a product of flat directions of
the
following schematic form:
\beq
\prod_{i=1}^n (\psi \psi \chi)^2_i, \ \ {\rm where} \ \ n=
\{1,\cdots ,4\} \label{generic}
\eeq
where each  factor is a  sixth order polynomial
analogous to (\ref{basic}),
we have suppressed the third family index, and each
factor  has a non-zero VEV
$v_i$ for each field in
$\psi_{2i-1}^{2i-1}
\psi_{2i}^{2i}
\chi_{2i-1,2i}$. (These directions are always F-flat.) For example,
one can consider 
\beq
(\psi^{3 1}_1 \psi^{3 2}_2 \chi^{3}_{[1,2]})^2 
(\psi^{3 3}_3 \psi^{3 4}_4 \chi^{3}_{[3,4]})^2 \label{example}
\eeq
for the case $n =2$, where the two sets of fields have VEV's $v_1$ and $v_2$,
respectively. The direction in (\ref{generic}) has the generic
surviving gauge group
\beq
U(12-2n)\times SO(8-2n) \times U(1)^n\; .
\eeq

This flat direction provides a field theory  realization
of a motion  of $n$ sets of (six) $D7$ branes
 in, say, the third complex plane, 
away from the fixed point at the origin, and  with each set at a different
location. The  T-dual
string theory construction  in terms of a generic
continuous Wilson line,
given by  (\ref{WL}), was derived in the previous Section. However,
note again that the field theory allows for a realization of  such
a motion of branes  only in the neighborhood of the original fixed point,
i.e., the result
is valid only in the power series expansion in terms of the VEV's of the
fields.

 When $p$ of the VEV's
in (\ref{generic}) are equal\footnote{One
can show that the enhanced symmetry also holds if each pair of
VEV's differs by a factor of $\omega$ or $\omega^2$. For example,
in (\ref{example}) the conserved $U(2)$ generators become
$F_{12}$, $F_{34}$, $\tilde{F}_{13}+\tilde{F}_{24}$,
and $\tilde{F}_{14}-\tilde{F}_{23}$,
where
$\tilde{F}_{13} \equiv i\left(\frac{v_1}{v_2}T^1_3 - \frac{v_2}{v_1}T^3_1 +
A_{13}\right)$, ${{v_1}\over {v_2}} = \omega$ or $\omega^2$, and similarly
for 
$\tilde{F}_{24}$. 
We have not found an analog of this freedom  in the continuous Wilson line
construction.},
the gauge factor $U(1)^p$ is enhanced to $U(p)$.
For example, if $v_1= v_2$ in (\ref{example}), it is straightforward
to show that the generators $F_{12}$, $F_{34}$, $F_{13}+F_{24}$,
and $F_{14}-F_{23}$ are conserved, and form the surviving group $U(2)$.
(The $U(1)$ generator is $F_{12}+F_{34}$.)
Most generally,  the direction
in which sets of $p_i$ of the $n$ VEV's are equal, with
$\sum_{i}p_i=n= \{1,\cdots,4\}$, leads to
\beq
U(12-2n)\times SO(8-2n) \times \prod_{i} U(p_i)\; ,
\eeq
in one to one correspondence with the continuous Wilson line solutions
(\ref{WL}) 
with  $p_i$  ($3\times 3$)-dimensional matrices $W_i$ equal, which 
in the T-dual picture realizes the 
the motion of 
$n$ sets of (six) $D7$ branes with groups of  $p_i$ of them  at
the same position, thus providing  for an enhancement of gauge symmetries.

One can generalize the construction of flat directions to include fields 
with all three family indices.  For example,
each factor in (\ref{generic}) can be generalized to a product of
three factors, one for each family with its own VEV $v^\alpha_i$
but the same gauge group structure. These directions are still
F-flat provided one does not mix families within the same
factor. These solutions provide a field theoretical realization of the 
motion of branes in multiple complex planes, whose T-dual string theory 
construction in terms of  multiple continuous
Wilson line solutions was given at the end of the previous
Section.

On the other hand, the  discrete Wilson lines
(\ref{wilson})
provided a
T-dual realization in terms of  $n$ sets of (six) $D7$ branes sitting
at a fixed  point   away from the
origin,  in, say,  the third complex plane; 
 the starting gauge group there is (\ref{discgg})
with the massless particle content (\ref{disc1})-(\ref{discmatter}).
The flat direction corresponding to the moving of one set of branes
away from the fixed point (located away from the origin)  is associated
with the HIP 
\beq
S^{\a{ i_3}}_{i_1} P^{\a{ i_1}}_{i_2} 
Q^{\a{ i_2}}_{i_3} \rightarrow
S^{31}_{1} P^{31}_{1} Q^{31}_{1},
\eeq
where the restriction to a single family, say, $\alpha=3$,  ensures 
F-flatness\footnote{Each factor $SPQ$ can again be replaced by a product of
three factors, one  with its own VEV for each family,
provided  the family indices are
not mixed within a factor. This corresponds to the multiple continuous
Wilson lines and in the T-dual picture to the motion of sets of branes in
three complex planes. }.
This breaks the $U(n)^3$ symmetry down to $U(1) \times U(n-1)^3$,
where the $U(1)$ generator is the diagonal sum  $t^1_1$ of three broken
$U(1)$ generators, where
\beq
t^i_j \equiv t^{(1)i}_j +t^{(2)i}_j+t^{(3)i}_j. \label{diagonal}
\eeq
In (\ref{diagonal}) $t^{(l)}, \; l=1 \dots 3$ is
the generator of the $l^{th}$ $U(n)$ factor.
Similarly, a product of $\left( SPQ\right)^{q}$
with distinct gauge indices, such as, e.g., 
\beq 
\left( S^{31}_{1} P^{31}_{1} Q^{31}_{1} \right)
\left( S^{32}_{2} P^{32}_{2} Q^{32}_{2} \right), \label{spq2}
\eeq
generically breaks $U(n)^3$ to $U(1)^q \times U(n-q)^3$.
However,  for special points with equal VEV's for each
of the factors\footnote{ 
There are also enhanced symmetries for
the products $\left( SPQ\right)^q$ in the case in which the VEV's of
each factor have relative phases of $\omega$ or $\omega^2$.
The generators of the enhanced symmetry are generalizations of
(\ref{diagonal}) in which there are corresponding phases in the
coefficients of the $t^{(l)}$. For example, if the two factors in (\ref{spq2})
have ${{v_1}\over{v_2}} = \omega$ or $\omega^2$, then 
$\tilde{t}^1_2 \equiv \frac{v_1}{v_2} t^{(1)1}_2 +t^{(2)1}_2+
\frac{v_2}{v_1}t^{(3)1}_2$,\ \ $\tilde{t}^2_1 \equiv \tilde{t}^{1\dagger}_2$,
$t^1_1$, and $t^2_2$ form a conserved $U(2)$. 
Again, we have not found an analog on
the string theory side. }
there is an enhanced symmetry
$U(q) \times U(n-q)^3$.

One can 
choose a ``hybrid''  flat direction composed of:
\beq
\prod_{i} (\psi \psi \chi)^{2 p_i}\prod_{j} (SPQ)^{q_j}
\ , \ \ {\rm where} \sum_{i}p_i\le 4-n\ ,\ \  \sum_{j}q_j\le n \; . 
\eeq
The surviving gauge group is 
\beq
U(12-2n-2\sum_ip_i)\times SO(8-2n-2\sum_ip_i)\times \prod_{i} U(p_i)\times 
\prod_{j} U(q_j)\times U(n-\sum_j q_j)^3.
\eeq

This field theory picture of course has an analogous Wilson line realization
encoded in a special choices   of $W_i$ matrices, including the choice of
the specific branches  for the interpretation of the phases $\phi_w$.

\section{Conclusions}

In this paper we focused on the study of continuous Wilson lines within
 four-dimensional N=1 Type IIB orientifold models.  We 
enforced unitarity and contructed the most general set
 of continuous Wilson lines  within the original  $Z_3$ orientifold
 \cite{ABPSS} and demonstrated that these models are in
 one-to-one correspondence
 with the T-dual models, where  each Wilson line  has an interpretation 
 of  $n$ sets ($n=1,\cdots,4$) of (six) branes moving in one (of the three)
  complex planes. 
 The number of parameters of such a continuous Wilson line is in one-to-one
 correspondence with the parameters that specify the location in the
complex
 plane  of each set of  branes. When  a sub-block of the 
 continuous Wilson line  becomes discrete, i.e., the sub-block that 
 commutes with the corresponding sub-block of the $\gamma_{Z_3}$ element and 
 depends on a discrete parameter, this corresponds in the 
  T-dual orientifold  to  branes sitting at the
$Z_3$ orbifold fixed points. The generic Wilson lines break the original
gauge group
$U(12)\times SO(8)$ down to $U(12-2n)\times SO(8-2n)\times U(1)^n$. 
 A gauge enhancement takes place for  special values of the Wilson
line parameters, e.g., when $k$ sub-blocks are equal then $U(1)^k$ is
promoted to $U(k)$. Similarly, when the full Wilson line becomes discrete
the gauge group is enhanced to 
$U(12-2n)\times SO(8-2n)\times U(n)^3$.  The Wilson line solution
continuously interpolates between the limit of no Wilson line
and the discrete solutions.

We  also analysed  the field theoretic analog, describing the above string
constructions  as  D- and F-flat deformations of  the effective field theory of
the original model as well as  deformations of the models with discrete Wilson
lines.   The field theory  describes these string solutions only in the
proximity of the original models, i.e., it allows  only  for a
power-series
 expansion in the vacuum expectation values of the chiral superfields, specified
 by the holomorphic gauge invariant polynomials that 
 parameterize the moduli space of the supersymmetric deformations of the
 original models. We find the explicit  form of the holomorphic polynomials,
 that are in one-to-one correspondence with the parameters of the  string
 constructions with the continuous Wilson lines (and their T-dual
 interpretation), thus quantifying the correspondence between the two
 complementary approaches. 
 
The work sets the stage for  further investigations of  models with continuous
Wilson lines. In particular, the explicit construction of the continuous Wilson
lines would allow one to construct not only the massless, but also the massive
spectrum of the string models. The dependence of the mass spectra
and the couplings on the  continuous Wilson line parameters, both from the
explicit string construction as well as from the (perturbative) field theory
perspective, deserves further study.

Another  more general  direction involves a study of a general class of
four-dimensional $N=1$ Type IIB orientifold models, in order to
establish the
general (and precise) correspondence  between the models with 
continuous Wilson lines, their T-dual
interpretation,  as well as their field theory realization \cite{CLPUW}.
In general these models  contain not only $D9$ branes but also, e.g.,
$D5$ branes. The latter  can be located at different points on a
particular two-torus $T^2$, where they are point-like,  thus allowing for
  even more
involved models,  implementing  simultaneously 
 moving branes {\it and} the  actions of continuous Wilson
lines. Investigation of these  general classes of string solutions (by
determining
the gauge group, the  mass  spectra and the  couplings)
 would shed light  on the  properties  of a broad class of
open string  models  within symmetric Type IIB orientifold constructions,
 and may in turn lead to a discovery of
  potentially realistic open-string solutions.

\acknowledgments

We would like to thank Angel Uranga  for a many
communications and suggestions  regarding the work presented in the paper as
well as for collaboration on related topics. 
We also   benefitted from discussions
with  L. Faccioli, S. Katz,  M. Pl\" umacher, and J. Wang. The work was
supported in part by
U.S.\ Department of Energy Grant No.~DOE-EY-76-02-3071 (M.C.\ and
P.L.) and  in part by the University
of Pennsylvania Research Foundation award (M.C.).

\clearpage

\def\B#1#2#3{\/ {\bf B#1} (19#2) #3}
\def\NPB#1#2#3{{\it Nucl.\ Phys.}\/ {\bf B#1} (19#2) #3}
\def\PLB#1#2#3{{\it Phys.\ Lett.}\/ {\bf B#1} (19#2) #3}
\def\PRD#1#2#3{{\it Phys.\ Rev.}\/ {\bf D#1} (19#2) #3}
\def\PRL#1#2#3{{\it Phys.\ Rev.\ Lett.}\/ {\bf #1} (19#2) #3}
\def\PRT#1#2#3{{\it Phys.\ Rep.}\/ {\bf#1} (19#2) #3}
\def\MODA#1#2#3{{\it Mod.\ Phys.\ Lett.}\/ {\bf A#1} (19#2) #3}
\def\IJMP#1#2#3{{\it Int.\ J.\ Mod.\ Phys.}\/ {\bf A#1} (19#2) #3}
\def\nuvc#1#2#3{{\it Nuovo Cimento}\/ {\bf #1A} (#2) #3}
\def\RPP#1#2#3{{\it Rept.\ Prog.\ Phys.}\/ {\bf #1} (19#2) #3}
\def\etal{{\it et al\/}}

\bibliographystyle{unsrt}

\end{document}